# Inhomogeneous Charge Distribution Across Gold Nanoclusters Measured by Scattered Low Energy Alkali Ions

Christopher Salvo, Prasanta Karmakar[1] and Jory Yarmoff*

*Department of Physics and Astronomy, University of California, Riverside, CA 92521*

**Abstract**

The neutralization of low energy $Na^+$ and $Li^+$ ions scattered from Au nanoclusters formed by deposition onto oxide surfaces decreases as the cluster size increases. An explanation for this behavior is provided here, which is based on the notion that the atoms in the clusters are not uniformly charged, but that the edge atoms are positively charged while the center atoms are nearly neutral. This leads to upward pointing dipoles at the edge atoms that increase the neutralization probability of alkali ions scattered from those atoms. As the clusters increase in size, the number of edge atoms relative to the number of center atoms decreases, so that that the average neutralization also decreases. Calculations employing this model are compared to experimental data and indicate good agreement if the strengths of the dipoles at the edge atoms are assumed to decrease with cluster size. This model also explains differences in the neutralization probabilities of scattered $Na^+$ and $Li^+$.

---

[1]Present address: Variable Energy Cyclotron Centre, 1/AF, Bidhannagar, Kolkata 700064, India

*Corresponding author, E-mail: yarmoff@ucr.edu

## I. Introduction

Over the past 30 years, a large body of research has been conducted to explore and explain the high catalytic activity of gold (Au) nanoclusters supported on oxide substrates [1-4]. These clusters are composed of several to hundreds of Au atoms and have reaction rates that rival those of enzymes, which is in contrast to bulk Au's extremely inert character. The activity depends on cluster size, with the highest rates for the oxidation of CO being those that are approximately 3.2 nm in diameter [5]. Nanoclusters have also found use in other applications such as functionalization capabilities in biology [6,7] and quantum computing [8]. Although it can be said that Au nanoclusters are the "gold standard" in nanocatalysts, there are a variety of other metals that also form catalytically active nanoclusters when deposited on an oxide support [1,9-12]. There are many ways to fabricate nanoclusters, including physical vapor deposition (PVD), chemical vapor deposition, buffer layer assisted growth, size selected deposition, and chemical synthesis [13-16].

PVD is one of the most popular methods used to grow Au nanoclusters on surfaces in ultra-high vacuum (UHV) [3]. Au atoms are deposited randomly onto a substrate, via a thermal atomic beam, on which they diffuse to spontaneously form the clusters. Au atoms deposited onto an oxide substrate follow a Volmer-Weber growth mode in which they initially coalesce to form flat, single atomic layer clusters and, with additional deposition, form multilayer clusters prior to deposition of the equivalent of a full monolayer [17]. Such clusters formed by direct deposition onto an oxide surface are flatter than gas phase nanoclusters that consist of the same number of atoms [18]. At each coverage of Au, there is a narrow distribution of nanocluster sizes on the substrate [19-21]. As more Au is deposited, the average size of that distribution increases until enough Au is present that the clusters combine into a complete thin film [17]. The detailed internal atomic structure of the supported nanoclusters is not well known, however, as techniques such as scanning electron or tunneling microscopy cannot image the individual atoms [17,22].

Multiple explanations of the high catalytic activity of Au nanoclusters have been proposed that include the roles of quantum size effects, increase of low-coordinated atoms, cluster morphology, substrate defects, and charge state [23-26]. Much of the current consensus is that the catalytic reactions occur at the edges of the nanoclusters via adsorption to the Au atoms that are directly bonded to oxygen in the substrate [27-29]. These edge atoms are presumably positively charged due to electron transfer in bonding to oxygen [30,31]. Note that a good deal of earlier work had suggested that Au nanoclusters were overall negatively charged [4,32-34], while other work has shown that the aggregate charge is positive [32,35-37]. Thus, the overall charge state of deposited nanoclusters is not firmly established and may depend on the particular materials involved.

Low energy ion scattering (LEIS) was initially developed in the 1960s as a means to provide information about the elemental composition of surfaces [38,39]. Most LEIS experiments employ noble gas ions as projectiles due to their ease of preparation and extreme surface sensitivity when using a charged particle detector, which is related to their irreversible Auger-type of neutralization [40].

In contrast, scattered alkali ions neutralize by a resonant charge exchange mechanism due to the overlap between their relatively small ionization potentials and the surface conduction bands [41]. Studies of scattered alkali ions show that the neutralization probability is dependent on the local electrostatic potential (LEP) just above the surface. There have been experimental and theoretical investigations of the neutralization probability in low energy alkali scattering from surfaces that have non-uniform local electrostatic potentials created by adsorbing small numbers of adatoms onto clean metal surfaces, which changes the overall work function, and then measuring different neutralization probabilities for scattering from each type of atom [42-46]. The results of these studies showed that the LEP close to the surface can vary leading to different neutralization probabilities for scattering from the different elements at the surface. In particular, scattering from a positively charged adatom, such as an alkali adsorbate at low coverage, has a significantly higher neutralization probability than scattering from the substrate due to the upward-pointing dipole that lowers the LEP above the adatom site [45-47].

Measurement of the neutralization probability of scattered low energy alkali projectiles is a useful tool for investigating deposited nanoclusters on oxides. This is because the alkali LEIS technique is adept for studying surfaces composed of multiple elements and can address questions about the charge state of those elements. Previous results have shown that the ions scattered from Au and Ag nanoclusters on an oxide and other substrates have a much higher neutralization probability than those scattered from the bulk metal, and that neutralization decreases as the cluster size increases [48-55]. Although it is possible that the alkali LEIS neutralization rate and the Au nanocluster catalytic activity are related, such a correlation has not yet been confirmed.

To explain the unusually high neutralization for alkali ions scattered from small Au nanoclusters, a model is developed here based on the notion that the charge associated with each Au atom in a cluster is different. Previously, the charge on the Au clusters had been considered in aggregate when analyzing ion neutralization [48-55], but the present model is dependent on the differences in charge between the atoms and not on the average charge. Density functional theory (DFT) calculations performed for Au nanoclusters supported on $TiO_2$ find that the charge associated with the edge atoms of the nanocluster are noticeably different from the center atoms [27,31]. Coincidentally, it is those same edge atoms that are the catalytically active sites [29,30,56]. A parametrized model is developed here to illustrate how the neutralization of singly scattered low energy alkali ions can depend on cluster size by combining the relative charge values of the center and edge atoms in a nanocluster, as determined from DFT, with the traditional paradigm for scattered low energy alkali neutralization. Inter-nanocluster interactions that affect the strength of the dipoles on the edge atoms when the clusters are close to each other also need to be included in the model to properly reproduce the experimental data. The success of this model shows how ion scattering can implicitly discern the presence of an inhomogeneous charge distribution within an individual deposited nanocluster. In addition, the neutralization probability differences between low energy $Na^+$ and $Li^+$ ions scattered from Au nanoclusters are consistent with this model [54].

## II. Experimental Procedure

Experiments are performed in an UHV chamber with a base pressure better than $5 \times 10^{-10}$ Torr. The sample is mounted on the foot of an XYZ rotary manipulator that allows it to access all of the sample preparation and surface analysis tools in the chamber. The tools include low energy electron diffraction (LEED), x-ray photoelectron spectroscopy (XPS), and low energy ion scattering (LEIS).

The substrate is a polished $TiO_2(110)$ single crystal purchased commercially. It is prepared in UHV by cycles of sputtering with 500 eV $Ar^+$ and annealing to 975 K for 15 min [49]. LEED and XPS are used to verify the crystal order and cleanliness. The nanoclusters are formed on the crystal surface by PVD using a thermal atomic beam of Au. The atomic beam is produced by running current through a W filament (Mathis) with Au wire wrapped around it that is mounted inside a Ta case with an aperture facing the sample [17,35]. The evaporation rate is calibrated by a quartz crystal microbalance (QCM).

LEIS is performed using time-of-flight (TOF) spectroscopy to measure both neutral and charged scattered particles. Beams of 2.0 keV $Li^+$ and $Na^+$ ions are produced simultaneously from thermionic emission guns (Kimball Physics) that are pulsed at 80 kHz, with pulse widths of roughly 100 ns. Once a projectile has scattered from the target, it travels down the 0.46 m long TOF leg where it is detected by a series of three microchannel plates (MCPs). The entrance to the first MCP is grounded so that neutral and charged particles are measured with equal sensitivity, although the absolute sensitivity decreases significantly with smaller scattered kinetic energies beginning at about 1.5 keV [57-59]. Each gun is separated from the TOF leg by 30°, leading to scattering angles of 150°. A set of deflection plates in the TOF leg is used to separate the charged and neutral particles. When both plates are grounded, all of the scattered particles pass through yielding the total counts, while placing 300 V across the plates deflects the ions allowing only neutral particles to be collected. Further details of the experimental procedure are found in Liu *et al.* [54].

The use of $Li^+$ and $Na^+$ ion beams incident on the sample simultaneously enables a comparison of data collected from the same surface with different projectiles. Since the projectiles have different masses but the same incident kinetic energy, they travel at different velocities. The difference in arrival times allows for the separation of features due to scattered Na and Li within a single spectrum.

In addition, the protocol is to never allow more than 1% of the surface atoms to be impacted before re-preparing the sample to minimize beam damage so that the data always reflect the intact clusters. This is easier when using TOF methods, as opposed to electrostatic analysis, because the pulsed beams have a very small duty cycle even with two ion beams impacting the sample.

**III. Experimental Results**

The LEIS technique is typically employed for the determination of the elemental composition of the first few atomic layers of a sample [39]. When TOF is used to measure all scattered particles, LEIS is surface sensitive because of shadowing and blocking. Shadowing occurs when atoms in the outermost layers prevent the incident ion beam from reaching the deeper layers so that direct scattering cannot occur. Blocking occurs when projectiles scatter from a second or deeper layer atom but are prevented from reaching the detector because of scattering by a surface atom. Shadowing and blocking are often illustrated by calculating the shape of cones, which are the regions behind each atom from which the projectiles are excluded. Because of shadowing and blocking, the projectiles that are detected are primarily those that scatter from the outermost atomic layers, making this technique ideal for studying materials such as catalysts where the chemistry takes place on the surface [60,61]. In analyzing LEIS data, it is assumed that the target atoms are located at specific lattice sites, but that they are unbound since the projectile energies far exceed the atomic bonding energies. It is furthermore assumed that the projectiles interact with one target atom at a time, which is the binary collision approximation (BCA).

Typical TOF-LEIS spectra for $Li^+$ and $Na^+$ scattered from Au nanoclusters are shown in Fig. 1 with the scattered particle yields on the y-axis and time on the x-axis. The x-axis is reversed so that higher energy scattered projectiles are to the right. The target is a well-ordered $TiO_2(110)$ substrate onto which 0.5 monolayers (ML) of Au are deposited, which is a small enough coverage that nanoclusters are formed rather than a Au film.

A single scattering peak (SSP) occurs when a projectile elastically scatters directly into the detector after making a hard collision primarily with one surface target atom. With the beam energy (2.0 keV) and scattering angle (150°) held constant, the kinetic energy of a singly scattered projectile is dependent primarily on the ratio of the target to projectile atomic mass [62]. Note that there is also a small amount of inelastic energy loss caused by the frictional forces of electrons in the sample that act on the projectile to broaden and lower the energy of a SSP, but the magnitude of this loss is typically on the order of 100 eV and can be ignored in the present analysis [63]. There are three clear SSPs present in both the total and neutral yield spectra shown in Fig. 1 due to the single isotope of Na and two isotopes of Li scattering from Au atoms. The kinetic energy of $^{23}$Na scattered from Au is 1290 eV and for $^{7}$Li it is 1750 eV (ignoring any inelastic energy losses), corresponding to flight times of 4.86 and 2.09 μs, respectively. The SSP for $^{6}$Li scattering from Au is also visible in the spectra, but its intensity is minimal due to its small isotopic abundance. Since Li is less massive than Na, it is scattered at a higher velocity, which leads to a shorter flight time. Fortunately, there is clear separation of the Na and Li signals so that these scattering events can be analyzed independently.

Multiple scattering occurs when a projectile interacts with more than one target atom, leading to a distribution of energies rather than to a sharply defined SSP. Multiply scattered Li particles are observed as a mound to the left of the Li SSP because Li is light enough to backscatter into the detector following multiple collisions with Au, Ti and/or O atoms and still retain enough kinetic energy to be detected. Fortunately, under the conditions of these measurements, the mound is sufficiently narrow that it is clearly separated from the Na SSP. There is no multiple scattering mound for scattered $^{23}$Na because it has a larger mass and thus cannot backscatter from the light substrate atoms.

Although Au is not the only element present on the surface and visible to the ion beams, SSPs due to scattering from Ti and O do not appear in the spectra. Na is less massive than Ti, but the scattered energy at 150° is 280 eV which is too low to be detected by the MCP [57]. Na is more massive than O and will thus not singly scatter at a large angle. Li can scatter from O but would have an energy of 350 eV, again being too low for detection. Li scatters from Ti with a kinetic energy of 1160 eV, or at a flight time of 2.57 µs, which can be observed by the MCP. The Ti signal is not large, however, because the Au nanoclusters cover part of the $TiO_2$ surface, the cross section for $^7Li$ scattering from Ti is approximately a factor of 3.4 smaller than for scattering from Au [62], and the Ti SSP is buried underneath the Li multiple scattering background.

When alkali ions scatter from a surface, their neutralization probability depends on the local electrostatic potential (LEP) above the target atom, the energy and width of the ionization $s$ level at a certain distance above the surface and the velocity of the projectile along its outgoing trajectory. The model most often used to describe this interaction is resonant charge transfer (RCT), which was originally developed for alkalis interacting with clean metal surfaces [41,64]. In the RCT model, the alkali's ionization level sees its image charge in the conductive metal substrate and shifts up in energy when it approaches the surface. Simultaneously, the ionization level hybridizes with the levels in the surface, causing it to broaden. Once the projectile is close enough to the surface, charge quantum mechanically tunnels between the target conduction band and the broadened and shifted $s$ level because they overlap due to the small ionization energies of alkalis. After the projectile has scattered and is sufficiently far from the surface along its exit trajectory, tunneling can no longer occur. In the limit of small velocity, the process would be adiabatic and produce 100% neutralization since the ionization levels of Li and Na, which are the $2s$ and $3s$ levels, respectively, both lie below the Fermi energy. Because the projectile velocities in the low energy regime are large on the scale of the electron tunneling rates, however, the charge transfer process occurs non-adiabatically, which leads to a measured neutralization probability that is frozen in while the projectile is still close to the surface, typically a few angstroms above the scattering site [64]. Although such freezing does not actually occur at a specific distance, as the interaction weakens exponentially, the freezing distance is defined as the effective distance above the scattering site at which the overlap of the Fermi energy and the broadened and shifted ionization level leads to the measured neutralization probability.

A clean metal surface is a simple case for ion neutralization because of the uniform lateral potential that allows the global work function to determine the neutral fraction (NF), but the process is more complex in the presence of an adsorbate. Because an adsorbate can lead to a non-uniform LEP, the neutralization would then depend on the LEP just above the scattering site, rather than on the global work function [41,65]. In this context, the LEP is sometimes referred to as the local work function. For example, if an electropositive adatom, such as an alkali, is adsorbed on a surface, it donates most of its outer shell electron to the surface creating an upward pointing surface dipole at the adatom site [66]. The dipole is formed by the positive charge of the alkali adatom and its negative image charge in the substrate. This upward pointing dipole reduces the LEP directly above the adsorbate which, in turn, increases the NF of a low energy alkali ion scattered from that adatom. At low alkali adatom coverages, where there are isolated non-interacting dipoles, the NF in scattering from the adatoms is larger than from bare areas of the surface, indicating that the LEP of the surface is inhomogeneous [46,47,65,67].

The neutralization probability is measured experimentally by integrating the neutral and total yield SSPs and taking their ratio. Before integration, the background of multiply scattered projectiles is first subtracted from the SSP. The background is estimated by fitting the region surrounding each SSP to a linear function. The error bars of each SSP area are estimated by taking the square root of the total number of counts, including the background counts, which is assuming that the error is purely statistical. This error is then propagated to determine the statistical error associated with each NF.

Figure 2 presents the NF of $Li^+$ singly scattered from Au on the $TiO_2$ substrate shown as a function of the average Au nanocluster size, while Fig. 3 shows the NF for scattered $Na^+$. The experimental data are shown by the solid circles. The average diameter of the nanoclusters was determined by calibration to the scanning tunneling microscopy (STM) measurements of Lai *et al.* using the amount of Au deposited as determined by the QCM [17]. It is seen that the NF is on the order of 30% for Li and 50% for Na scattered from the smallest Au nanoclusters produced here, and that it decreases with cluster size until it reaches the same value as that for scattering from bulk Au, as reported previously [48,49,54]. Note that the figures show experimental data as well as simulations generated by the model described in the following section.

**IV. Model**

The model developed here to explain the change of NF as a function of cluster size is based on the notion that all of the atoms in a nanocluster are not electrostatically equivalent so that there is a difference in the neutralization probability of a scattered alkali projectile depending on the particular atom that is impacted. This assertion is supported by DFT calculations that show that the edge atoms of Au clusters deposited onto $TiO_2$ are positively charged, with an average Bader charge per atom of about $+0.0458e$, while the center atoms are nearly neutral [27,31]. This is illustrated schematically in the insets to Fig. 4, where the edge atoms are labeled with a "+" sign, while the neutral center atoms are unlabeled. Photoelectron spectroscopy measurements have suggested that Au nanoclusters have some positive charge, which supports the results of the DFT calculations, but they were not able to distinguish differences in charge across a nanocluster [32,35]. The positively charged atoms at the edge form upward-pointing dipoles leading to a higher LEP above them than the LEP above the nearly neutral interior atoms so that alkali ions scattered from the edge atoms of the nanocluster would have a higher neutralization.

The clusters are modeled here as being flat and consisting of a single atomic layer with a fcc hexagonal packing pattern and a (111) surface orientation. The distance between the Au atoms is set to 0.408 nm, which is the interatomic spacing in bulk Au metal. Single layer cluster sizes are employed that contain 7 to 217 atoms, which correspond to diameters from 0.82 through 6.5 nm. Although no crystalline structure is observed for actual deposited Au nanoclusters, this structure is chosen for the model because the (111) face is the low energy surface of fcc metals [17]. Some examples of these model nanoclusters are shown as insets in Fig. 4. In reality, the actual nanoclusters are not hexagonal (this would have been observed with STM, high resolution scanning electron microscopy, or transmission electron microscopy [17,68,69]) and the distance between the Au atoms can change depending on the cluster size and shape and the substrate material [17,70]. Nevertheless, the smallest Au clusters on surfaces are flat and one atomic layer thick, so that this approximation is sufficient to illustrate the basic physics of how the NF depends on cluster size.

As the size of the clusters increases, the number of edge atoms relative to center atoms decreases so that there are relatively fewer positively charged target atoms. Thus, the overall NF should decrease as the cluster size increases. Figure 4 shows the ratio of edge to center atoms using this approximation for the cluster atomic structure, and the ratio does go down as the clusters get larger. A numerical model is presented here to show that this idea can reproduce the observed NF vs. cluster size data.

A simple way to calculate the expected NF for Na$^+$ and Li$^+$ scattered from Au nanoclusters using this model is to assign different neutralization probabilities to the edge ($NF_E$) and center ($NF_C$) atoms in each cluster. It is further assumed that the LEP above the interior atoms is the same as that of neutral bulk Au, which leads to neutralization probabilities of $NF_C$ = 3% for scattered Na$^+$ and $NF_C$ = 9% for Li$^+$, as determined experimentally [49,71]. Since the edge atoms are positively charged, scattering from them would produce larger neutralization values. $NF_E$, the neutralization probability in scattering from the edge atoms, is thus a parameter in the model that needs to be determined. The overall NF generated by the model for a particular cluster size is calculated by averaging the individual neutralization probabilities in scattering from each of the Au atoms in the simulated clusters as

$$\overline{NF} = \frac{(N_E \times NF_E) + (N_C \times NF_C)}{N_E + N_C} \quad (1)$$

where $N_E$ and $N_C$ are the number of edge and center atoms, respectively, for a given nanocluster size.

If the atoms in the nanoclusters were in the form of Au(111) fcc crystallites, the normally incident ion beam would be able to impact the first three atomic layers, as any deeper layers would be shadowed. When the clusters are considered to be three atomic layer thick fcc crystallites and the highest possible value of $NF_E$ = 100% is used along with $NF_C$ = 3% for all the other atoms in the top three layers, the overall NF at a given Au coverage is much lower than the experimental values for a large range of cluster sizes when using Na$^+$ projectiles. Even assuming that the ions probe only the top two atomic layers produces NFs that are smaller than the experimental data. Assuming that scattering only occurs from the outermost Au layer, however, generates NFs that better match the experimental data. This was similarly tested for using Li$^+$ by setting *NF$_C$* = 9% and showing that if scattering from second and third layer atoms is included, there is no value for *NF$_E$* that will fit a majority of the data. At best, with *NF$_E$* set to be near 100%, only a few mid-range cluster sizes would match the data and the majority of cluster sizes would have an overall NF that is too high. If a smaller value is used for *NF$_E$*, only the smallest cluster sizes would match the data until a bulk film is formed.

For the above reasons, the model considers the clusters to consist of only a single atomic layer. This could mean that either the actual clusters are all just a single atomic layer thick, which is not consistent with STM data [9], or more likely that the clusters are not crystalline but instead are more amorphous structures in which the second layer and deeper-lying Au atoms are all shadowed or blocked by the outermost Au atoms and do not contribute to the SSP. Shadowing and blocking during scattering from such amorphous clusters is likely as their actual structure is sufficiently different from Au(111) and Au is a heavy atom that produces large shadow and blocking cones. The shadow cone radii of 2.0 keV Li$^+$ and Na$^+$ ions scattered from Au, for example, are calculated to be 1.05 and 1.25 Å, respectively, at a distance of 4 Å beyond the Au atom [72], which is close to the spacing between Au atoms. Since the shadow cone radii for Li$^+$ and Na$^+$ are not too dissimilar, it is likely that both projectiles probe approximately the same depth. The agreement between the model and experimental data, discussed below, shows that the use of single-layer hexagonal clusters provides a reasonable approximation to the actual structures of the clusters for the purposes of the model.

One strategy to estimate the neutralization probability of Na$^+$ and Li$^+$ ions scattered from the edge atoms, $NF_E$, is to calculate the dipole moment and then compare it to experimental NF data from another system with a known dipole moment. The dipole at the edge atoms is calculated by multiplying the distance $d$, between the two charges, by the absolute charge $q$ of one of them. The charge for the edge atom is taken from the average of all of the edge atoms' charge as calculated in Ref. [27]. The distance is estimated to be two times the bond length of Au-O, which is determined by first-principle calculations to be 2.08 Å [73]. This is the classic image charge problem, which generates an average dipole value for edge atoms directly attached to a substrate oxygen atom of 18.0±2.7 D. Reference [27] also provides values for the charge on the interior atoms and, in a similar manner, these values produce an average dipole moment of 1.4±1.6 D.

A comparison to previously published data is used to relate the dipole moment at an adatom site to the NF. Weare and Yarmoff determined that for very low coverages of Cs adatoms on Al(100), a dipole moment of 15.2 D is generated by the Cs adatoms [47]. Using 2.0 keV Li$^+$ projectiles (same as in the present experiment), a NF of ~70% was observed in scattering from Cs. Therefore, the value for Li scattering from the edge atoms based solely on the dipole strength would be a bit more than 70%. Although the relationship between dipole moment and NF is not linear, a simple extrapolation would predict a $NF_E$ of 81%. Since Cs and Au are relatively close in mass, any changes in neutralization due to differences in their scattered energy are minimal. Note that experimental data for the neutralization of Na$^+$ scattered from Cs adatoms are not available, so a similar estimate cannot be made.

If a value of 81% is used for the Li$^+$ $NF_E$, however, the model does not numerically match the experimental data, although it does have the same general trend. The fact that such an analysis of dipole strength does not work to precisely determine a neutralization probability is either due to inaccuracies in the estimate of the dipole strength or because there is a distinct difference between edge atoms of nanoclusters on $TiO_2(110)$ and isolated Cs adatoms on Al(100). One of these differences could involve effects of the substrate on the broadening, shifting, and freezing distance, as these could be considerably different for a metal and insulator. Therefore, the neutralization probability in scattering from a positively charged Au adatom in a nanocluster is not necessarily the same as the NF in scattering from an alkali adatom on a metal.

Since it is not possible to estimate the neutralization probability for scattering from the edge atoms directly from the dipole strength, a simple model is instead developed in which a best fit of $NF_E$ to the experimental data is performed. The results of this simulation for $Li^+$ scattering from Au clusters using $NF_E = 31\%$ and $NF_C = 9\%$ are shown in Fig. 2 by the dashed-dotted line labeled "Simple Model". Correspondingly, the results of a simulation for the scattering of $Na^+$ in which $NF_E = 63\%$ and $NF_C = 3\%$ are shown in Fig. 3. Both of these simulations match the experimental data fairly well for cluster sizes up to about 3.7 nm, but the simulated NF for larger clusters is too high for both projectiles. The fact that the model shows a decrease in NF with cluster size and the values agree with the experimental data for both projectiles in the same nanocluster size region indicates that the underlying physics of the model is basically sound, at least for the smaller clusters.

The failure of the simulations for larger cluster sizes is likely caused by a change in dipole strength with cluster size, which would thereby affect the LEP and thus the neutralization probability in scattering from edge atoms. There are two possible ways in which this could be explained. First, the charge associated with the edge atoms could depend on the cluster size as a fundamental property of the clusters. Second, as the nanoclusters grow in size they become more densely packed and are thus closer together. At a certain coverage, the edge atoms would be close enough to those of neighboring clusters to interact and depolarize the dipoles so that their strengths reduce, thus increasing the LEP above the edge atoms and decreasing $NF_E$. A similar effect is seen for high coverages of alkali adatoms on metals surfaces in which the dipole strength reduces when the alkali coverage reaches the point at which the adatoms interact with each other [47,65,67].

There is no simple physical reason to suggest that the edge atoms' charge would depend on the cluster size independent of neighboring clusters, but it is not impossible. The data suggest otherwise, however, since a constant value of $NF_E$ works for projectiles scattered from nanoclusters less than about 3.0 nm in diameter. If the edge atom charge were size dependent, it would be unlikely that this dependency would not affect the dipoles at the edge atoms of the smaller clusters.

Thus, the second idea is used to develop a modified model that incorporates the effects of a reducing dipole into the simulation as a consequence of interactions between clusters. In this more complete model, the effective neutralization of alkali ions scattered from the edge atoms, $NF_{EE}$, is adjusted as a function of cluster size to include the distance between clusters and $NF_E$ is considered to be a constant that represents the neutralization probability in scattering from edge atoms in non-interacting clusters.

To use this idea to produce a modified mathematical model that works for any cluster size, an equation is developed that depends on the distance between clusters, specifically the distance between the edge atoms of the nearest clusters, $d$. The method adjusts $NF_{EE}$ as a function of $d$ with the aid of a fitting parameter. The average distance between the edges is calculated by taking the square root of the inverse density of clusters and subtracting the average cluster diameter, both of which are obtained from Lai *et al.* [17]. This generates the average distance between the edges of nearest neighbor clusters $d$ in units of nm, which is plotted by the squares in Fig. 4. The equation used to determine $NF_{EE}$ is

$$NF_{EE} = (NF_E - NF_C)e^{-A/d} + NF_C, \qquad (2)$$

where A is the fitting parameter. An exponential dependence on distance is a reasonable approximation of the electrostatic interaction between nearby dipoles. The equation reduces to the correct values at the limits when they are infinitely spaced ($d = \infty$) and when the nanoclusters are touching ($d = 0$). For infinitely spaced nanoclusters, there is no interaction between nanoclusters so that the neutralization in scattering from the edge atoms $NF_E$ is the same as that determined from the simple model. For clusters that are so close that their edge atoms are adjacent, they are all now essentially center atoms so that the neutralization probability would be equal to $NF_C$.

The value of A that produces a best fit to the data is found to be 0.55 nm for both Li and Na, implying that this effect is related to the nanoclusters and not to the projectiles. The value of A is also relatively close to the dipole length of 0.42 nm implying that it is physically reasonable. $NF_E$ for both projectiles are slightly adjusted from the original estimates to $NF_E$ = 33% and $NF_E$ = 66% for $Li^+$ and $Na^+$, respectively, to better match the experimental data. These neutralization probabilities are slightly higher than those of the simple model, which is reasonable since it did not take the inter-cluster effects into account so that the best fit resulted in lower values.

Since the dipole-dipole interactions in this modified model are similar to the behavior of alkalis adsorbed on metal surfaces, the distance at which the alkali adatoms begin to interact is compared to the value of $d$ at which the NF in scattering from nanoclusters is no longer consistent with the simple model. Measurements of the work function as a function of coverage for alkalis deposited on metals initially decrease with coverage and show a minimum at the point at which the adatoms begin to interact, which occurs at an adatom-adatom spacing of roughly 0.91 nm for Na on Al(100) [74]. The average size of the nanoclusters beyond which the simple model no longer agrees with the experimental data is around 3.9 nm, which corresponds to an average spacing of $d$ = 1.4 nm. Note that this distance between clusters was calculated for the hexagonal clusters employed in the model, but if the same calculation is performed to find the average distance between the edges of circular clusters, the distance is 0.89 nm for 3.9 nm nanoclusters, which is essentially the same as the distance at which alkali adsorbates begin to interact. The data thus support the notion that interactions between nanoclusters are responsible for the reduction in the effective neutralization probability of projectiles scattered from the edge atoms for the larger cluster sizes.

The results of this modified model are shown in Figs. 2 and 3 as solid lines labeled "Modified Model", and they reproduce the experimental data fairly well over the complete range of cluster sizes. Above a cluster size of 4.5 nm, the clusters begin to agglomerate to eventually form a Au thin film [17] so that the ratio of edge atoms to center atoms and the value of $d$ both go to zero, as seen in Fig. 4. This is the cause of the kink in the modified model at 5.4 nm, as no separated clusters remain at that point. Hence, in the limit of large clusters, the NFs of the scattered ions reach the values associated with bulk Au.

Note that the smallest clusters that are considered in the model consist of seven Au atoms, with six of them being edge atoms. It is possible to have even smaller clusters in which there are no center atoms, with the limit being a single Au adatom. It is shown in Ref. [36] that the charge on a single Au adatom on $TiO_2$ is $0.2e^-$, which is less than the charge on the nanocluster edge atoms calculated by DFT [27]. Thus, the NF in scattering from a single Au adatom would be less than that of the smallest clusters considered here. This implies the possibility that the NF could drop for extremely small cluster sizes. Such a behavior has been observed in low energy alkali ion scattering from Au and Ag nanoclusters in Refs. [48,50,51], and a decrease of edge atom charge for clusters that are smaller than those considered here may be the reason. Thus, the extrapolation of the model to a cluster diameter of zero in Figs. 2 and 3 may not be borne out by experiments with smaller clusters that approach the limit of a single isolated Au adatom.

## V. Discussion

The model presented here is based on assigning different neutralization probabilities for scattering from edge and center atoms and showing that such differences can explain why the overall NF decreases with cluster size. It uses values for $NF_C$ measured from bulk Au but does not independently determine the $NF_{EE}$ probabilities from experiments, first principle calculations or other methods. Nevertheless, the numbers that are fit to the data do provide a good match between the model and the experiment. These values are now examined to determine if they are reasonable within the context of the general RCT process.

Figure 5 is a schematic diagram, used as an aid in this endeavor, which graphically shows how the energy levels of the projectiles are modified in proximity to the substrate. The horizontal axis is Z, the distance between the projectile and the surface, and the vertical axis is energy, E, where $E_{vac}$ is the vacuum level. Both the Li and Na $s$ levels are shown on the right side at their respective ionization energies as though they are infinitely far away and thus very sharp (not drawn to scale). As the projectiles get closer to the surface, they shift up in energy and broaden; this process is represented by curved dotted lines with arrows at the end that show the sharp $s$ levels becoming broad Gaussian functions at their respective freezing distances (numerical values discussed below) [41]. To the left of the energy axis are colored areas representing filled levels in the surface, although the diagram does not indicate the actual density of states of those levels. $E_{Fcenter}$ and $E_{Fedge}$ are the energies of the highest occupied levels for the center and edge atoms of the nanoclusters, respectively, which are not the same because of their different LEPs. The neutralization probability for a given scattering event would be the fraction of the broadened ionization level below the highest filled level (indicated by the horizontal dashed lines) at the freezing distance. An important thing to recognize is that magnitudes of both the shifting and the broadening can alter the neutralization probability in different directions. The freezing distances, shifting and broadening are now estimated to see if the numbers used for the neutralization of $Li^+$ and $Na^+$ scattered from edge atoms in the model are reasonable.

The freezing distance of the projectile is the effective maximum distance above the surface at which electrons can tunnel in and out of the ionization level. An equation from Kimmel and Cooper is used to estimate the freezing distances, which was derived by assuming that the coupling between the atom and the metal decays exponentially as the atom-surface separation grows [64]. The equation contains parameters that are dependent on the particular projectile and target used. The most relevant parameter is the perpendicular component of the velocity of the scattered particle, which is calculated classically for Li and Na singly scattered from Au [39]. Another parameter is the level's half width, which depends on the particular target and projectile, and was determined for the Li 2$s$ level in scattering from Cu(001) by Onufriev and Marston [75]. To the authors' knowledge, there are no explicit calculations for the Na 3$s$ level, so that the values for Li are used to generate freezing distances for 2.0 keV Li and Na scattered from Au of 3.7 and 4.2 Å, respectively. Furthermore, it is assumed that the freezing distance for a given projectile is the same when scattering from an edge or center atom. These distances are indicated on the horizontal Z-axis in Fig. 5. The accuracy of these values is limited by the above-mentioned assumptions used in calculating them.

Since the freezing distance of Li is smaller than that of Na, it is possible to create a scenario that satisfies the neutralization probabilities for the two alkali ions when scattering from edge atoms in the Au nanoclusters. The purpose here, however, is not necessarily to determine the absolute value of each parameter, but to show that reasonable values can reproduce the experimental results since an inaccuracy in one number can be compensated for by changing another. For example, it is assumed in this scenario that the LEP is independent of freezing distance, as the change of LEP with Z is unknown. Although it is quite likely that the LEP is not the same at the two freezing distances, the effect of including actual LEPs can be compensated for by altering the assumed positions of $E_{Fedge}$ and $E_{Fcenter}$ to produce the same NFs.

The Li 2$s$ level is assumed to have a width of 0.14 eV at its freezing distance, while the Na 3s has a width of 0.049 eV. In addition, it is assumed that the Li 2s shifts up 0.10 eV more than Na 3s. This leads to values of $NF_E$ equal to 33% for Li and 66% for Na, which are the same as those found by fitting the model to the experimental data. The neutralization values in scattering from the edge atoms are obtained by calculating the proportion of the Gaussian broadened ionization levels at the freezing distances that are below the effective Fermi energy indicated by $E_{Fedge}$. Similarly, the neutralization probabilities for Li and Na scattered from the center atoms are calculated using their highest filled energy, as defined by $E_{Fcenter}$. To match the NF values for ions scattered from bulk Au, which is equivalent to scattering from center atoms, $E_{Fcenter}$ must be 0.11 eV lower than $E_{Fedge}$. This generates $NF_C$ = 9% and 3% for Li and Na, respectively, consistent with the experimental values for scattering from bulk Au that are used in the model.

Within the formalization of the neutralization model presented here and the standard RCT process, and using the above values for the freezing distances, the difference in the NFs of scattered Li and Na for the smaller nanoclusters are rationalized. To see if the values used for broadening and shifting of the Li and Na *s* levels are reasonable, they are compared to numbers obtained from theory in the literature. Nordlander and Tully calculated the shifting and broadening for Li 2*s* and Na 3*s* when scattering from a jellium surface [76]. For Li and Na at their respective freezing distances, the calculations indicate that the broadening of the Li 2*s* and the Na 3*s* levels are 0.136 and 0.054 eV, respectively. These values are is in good agreement with the above scenario, especially considering the difference in substrates. Nordlander and Tully find that the width of the Li 2*s* level is a factor of 2.5 larger than the Na 3*s* at the freezing distances, while a Li 2*s* width that is a factor of 2.9 larger is needed to model the actual data. This ratio of the Li 2*s* and Na 3*s* widths is close, indicating that the model is physically realistic, although it should be noted that the ratio of the widths is not as important as their actual magnitudes.

Nordlander and Tully coincidently calculate the same value of 4.35 eV for the level shifting at the freezing distances for both the Li 2*s* and Na 3*s*, but this would not produce the observed differences in Li and Na neutralization. Therefore, there must be a small difference in the level shifting for Li and Na ions at their freezing distances. Since the Au nanocluster/oxide substrate target is much different than the Jellium surface used by Nordlander and Tully, it is not surprising that their calculations do not produce precisely correct values in this case, and the difference between the levels of 0.10 eV determined here by fitting to experimental data is relatively small. Also, considering the fact that Li has a smaller freezing distance and the shifting is modeled as an exponential, it is reasonable that the Li 2s would shift up a small amount more than Na 3s. Note that it is the difference between the levels that is important and the magnitude of that difference is implicitly set to match $NF_E$ and $NF_C$ for Li 2*s* and Na 3*s*.

Since the Li 2*s* level in the isolated atom has a lower energy than Na 3*s*, at first it may be expected that scattered Li should always have a higher NF than Na scattered from the same target. This is not the case for the small nanoclusters, however, although it is true for nanoclusters larger than 4.5 nm, as seen by comparing the experimental data in Figs. 2 and 3 and the data in Ref. [54]. Reference [54] did not, however, provide a quantitative explanation for why this is the case. To match the unexpected experimental data within the framework of the model presented here, Na must have a higher NF than Li when scattering from the edge atoms while Li must have a higher NF when scattering from the center atoms. Such relative neutralization probabilities are produced by the model presented here.

There are other processes that can affect the neutralization probabilities that are reported in the literature, but they would not alter the proposed model. First, a lowering of the ionization level energy is reported for projectile-target distances of less than 2.6 Å for Li scattering from highly oriented pyrolytic graphite (HOPG) [77], as opposed to the exponential upward shift used here [76]. It is calculated above that the freezing distance for Li is 3.7 Å, however, which indicates that even if such a downward level shifting did occur at those small distances, it would not alter the neutralization probability for this system. Second, it has been shown that the parallel component of the velocity of the exiting projectiles can have a significant impact on the measured neutralization probabilities [78,79]. Such parallel velocity effects are not important here, however, as the emission is along the surface normal.

**VI. Summary**

Au nanoclusters are a nanometer-sized form of solid-state matter with a non-uniform charge distribution. Since the measured NF of alkali LEIS is the average neutralization when scattering from all of the atoms in a nanocluster, differences in the LEP above the edge and center atoms due to the charge associated with the individual atoms must be considered in interpreting the data. A parametrized model is developed here, which is based on the notions that (1) the neutralization in scattering from edge atoms is larger than for scattering from center atoms and (2) that the ratio of edge to center atoms reduces with cluster size. The modified form of the model includes inter-nanocluster effects that act to depolarize the dipoles associated with the edge atoms when they are close to each other, and it is found to match the experimental data over all cluster sizes. This work shows how ion scattering is sensitive to the non-uniform charge distribution within a nanocluster.

With the use of the parametrized model presented here, the RCT formalization is applied to determine the level broadening, shifting, and freezing distance for Li and Na scattering from Au nanoclusters. Values for level broadening, shifting, and freezing distance are generated to verify that the neutralization probabilities calculated by the model for edge and center atoms are reasonable by comparing them to calculations [76]. These values also explain the unexpected larger neutralization probability of Na$^+$ compared to that of Li$^+$ when scattering from small Au nanoclusters.

A difference in the LEP between edge and center atoms in the nanoclusters is therefore implicitly identified by experiments that measure the neutralization of scattered alkali ions as a function of cluster size. These differences are consistent with the charges on the nanoclusters atoms calculated by DFT [27,31]. In comparison, STM has not been able to discern any differences in charge between atoms in a cluster partly due to the large size of the tip compared to the size of individual atoms. Kelvin force probe microscopy (KPFM) does directly measure the LEP and has seen differences at the edges of certain nanostructures [80], but its spatial resolution is even larger than that of STM. Furthermore, neither STM nor KPFM have the elemental resolution of LEIS.

Finally, this work suggests a correlation between the measured NFs and the enhanced catalytic activity of small Au nanoclusters. The measured NFs are explained by the increased positive charge of the edge atoms and how the ratio of edge to center atoms decreases with cluster size. It is also discussed above how extremely small clusters could have a reduced charge. Experiments of catalytic activity vs. cluster size show a maximum activity at a rather small cluster size [5]. Also, there is much evidence that the edge atoms are the active sites for catalysis [27,81-83]. Thus, it is not unreasonable to infer that both the catalytic activity and the neutralization of scattered low energy alkali ions depend on this same positive charge that is associated with the edge atoms.

## VII. Acknowledgements

The authors would like to thank Frank Liu for assistance with the data collection. This material is based upon work supported by the National Science Foundation under CHE - 1611563.

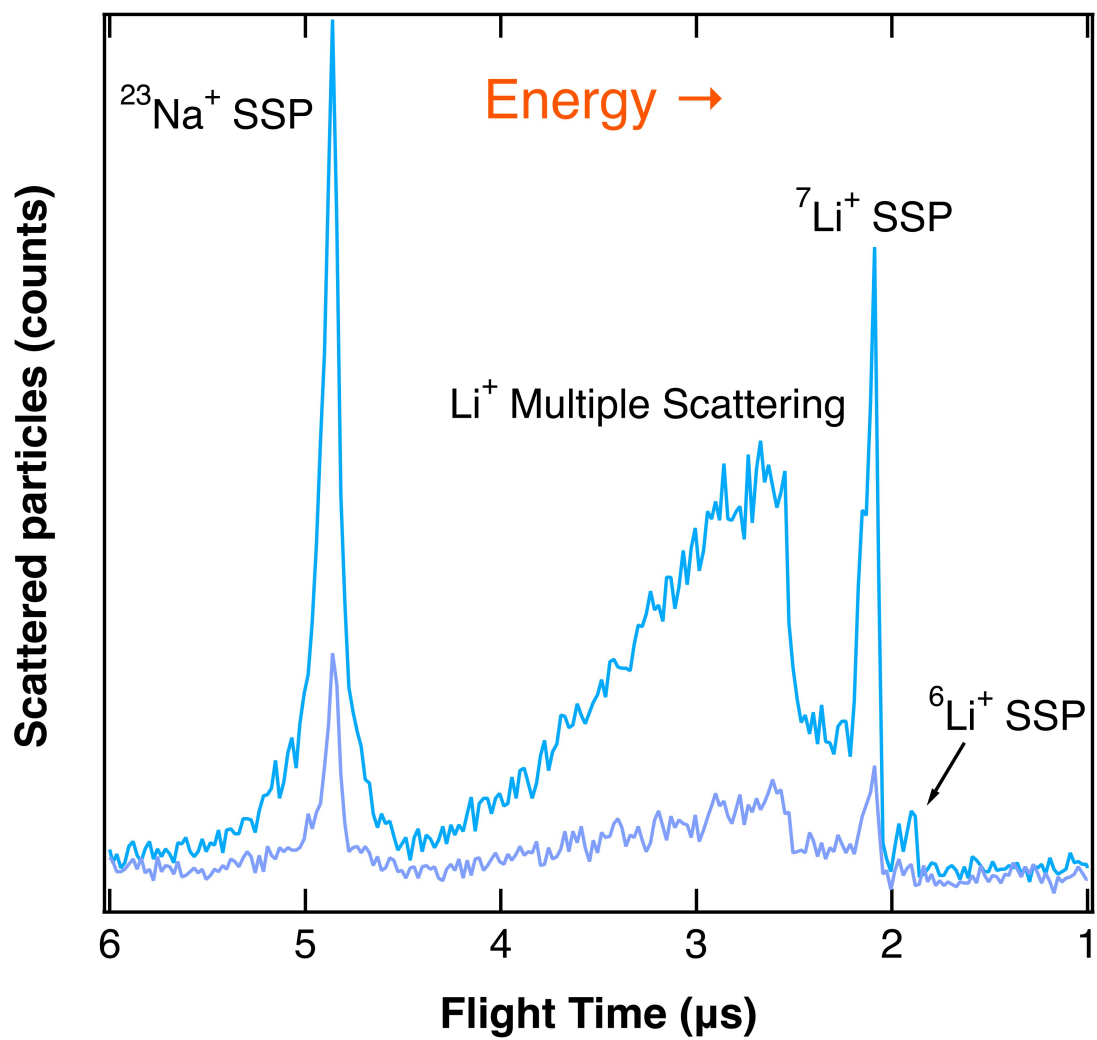

**Figure 1.** TOF-LEIS spectra of 2.0 keV Li$^+$ and Na$^+$ scattered from 0.5 ML of Au deposited on TiO$_2$(110) shown as a function of the flight time from the sample to the detector. The upper spectrum (blue) is the total yield and bottom spectrum (purple) is the neutral yield.

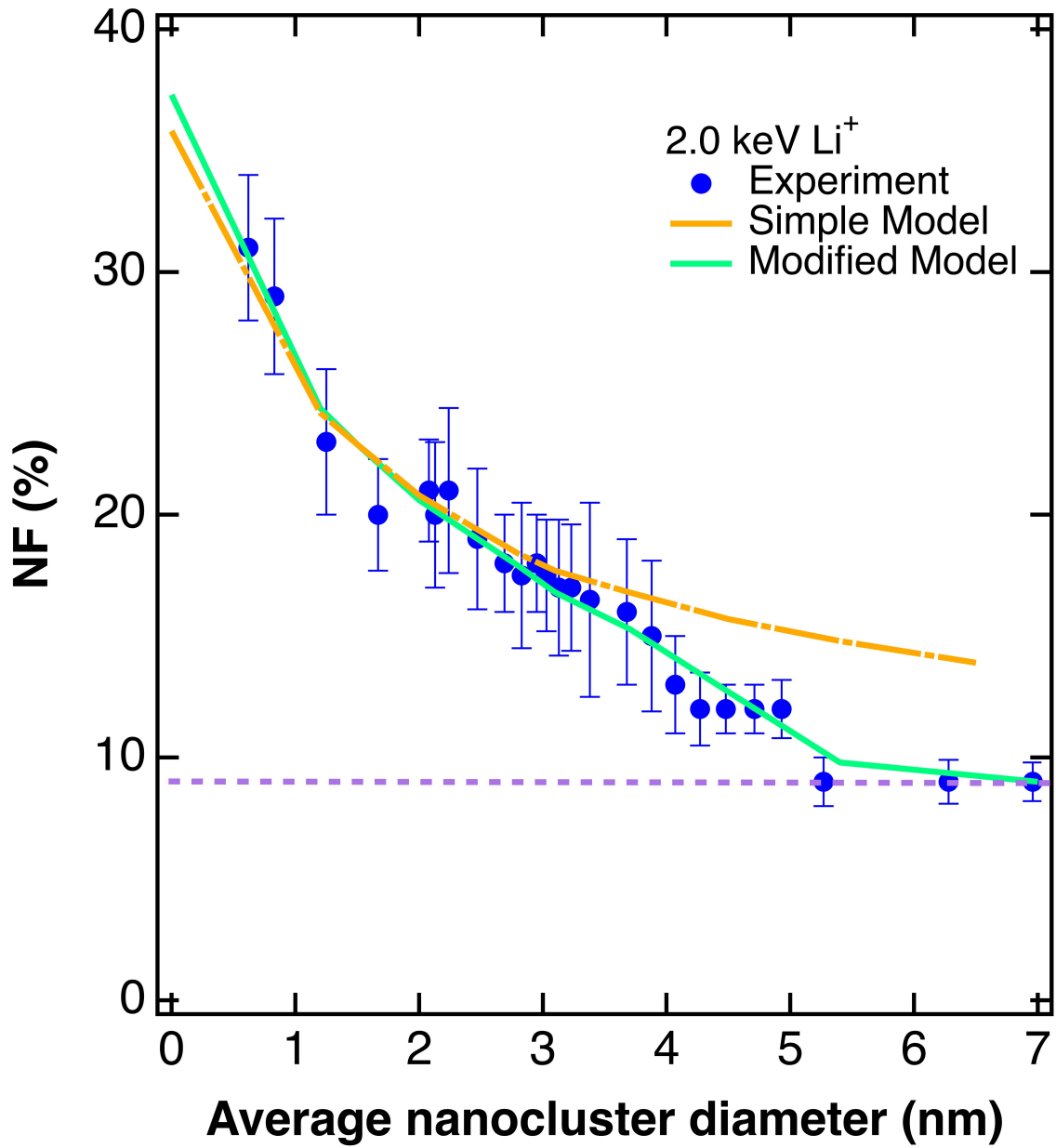

**Figure 2**. Experimental and simulated data of the NF of 2.0 keV Li$^+$ singly scattered from Au shown as a function of the average nanocluster diameter. The filled circles indicate the experimental data, the upper fragmented line (gold) is the simple model with $NF_E$ = 31%, and the solid line (green) is the modified model with $NF_E$ = 33% (see text). In all of the simulations, $NF_C$ is set to 9%, which is the value for scattering from bulk Au as represented by the horizontal dashed line.

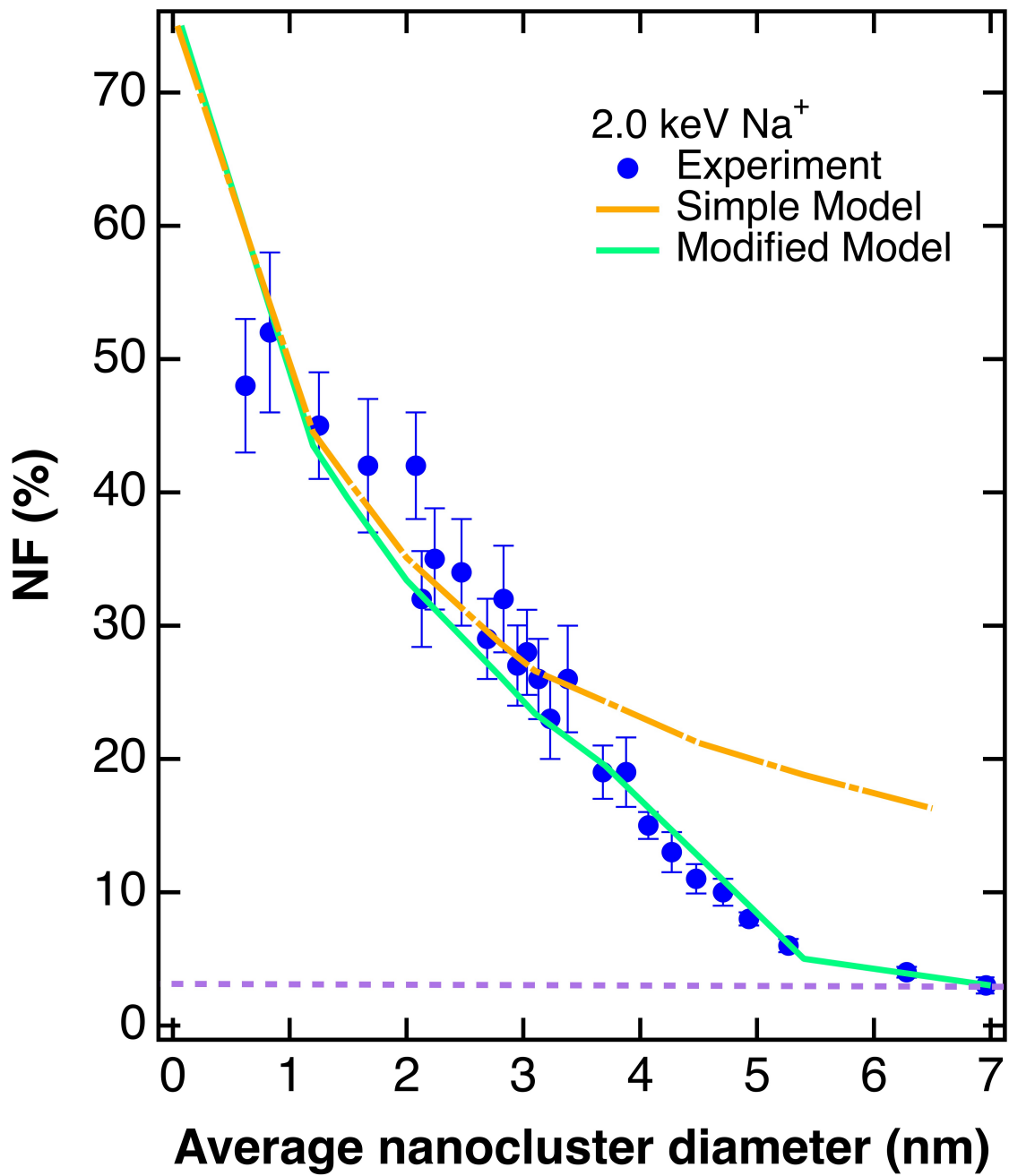

**Figure 3.** Experimental and simulated data of the NF of 2.0 keV Na$^+$ singly scattered from Au shown as a function of the average nanocluster diameter. The filled circles show the experimental data, the fragmented line (gold) is the simple model with $NF_E$ = 63%, and the solid line (green) is the modified model with $NF_E$ = 66% (see text). In all of the simulations, $NF_C$ is set to 3%, which is the value for scattering from bulk Au as represented by the horizontal dashed line.

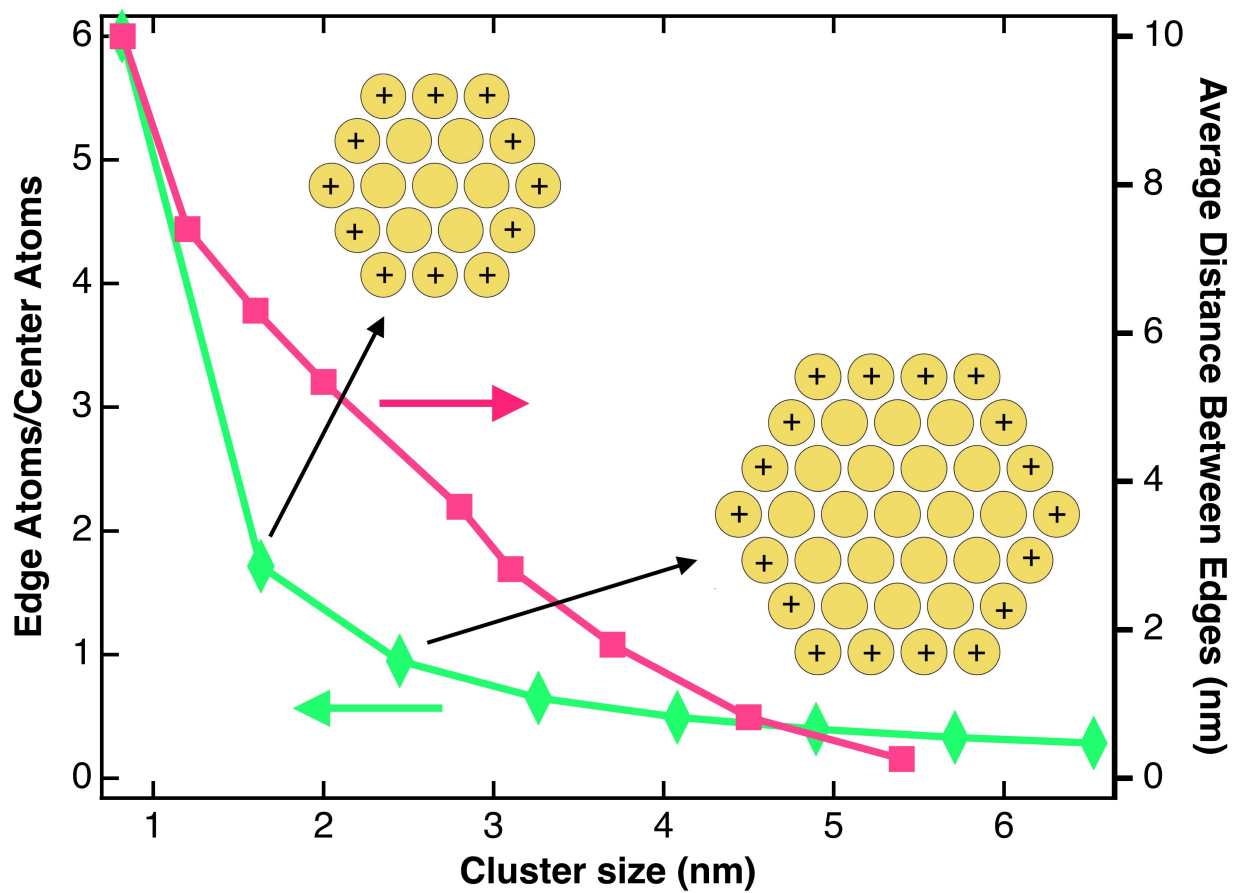

**Figure 4.** The ratio of edge atoms to center atoms as a function of cluster size used in the model is represented as diamonds with respect to the left axis (see text). The average distance between the edges of nanoclusters, *d*, is represented by squares with respect to the right axis. The insets show schematic diagrams of representative Au nanoclusters that correspond to two cluster sizes: 19 atoms with a diameter of 1.6 nm and 37 atoms with a diameter of 2.5 nm. The plus signs indicate the positively charged edge atoms.

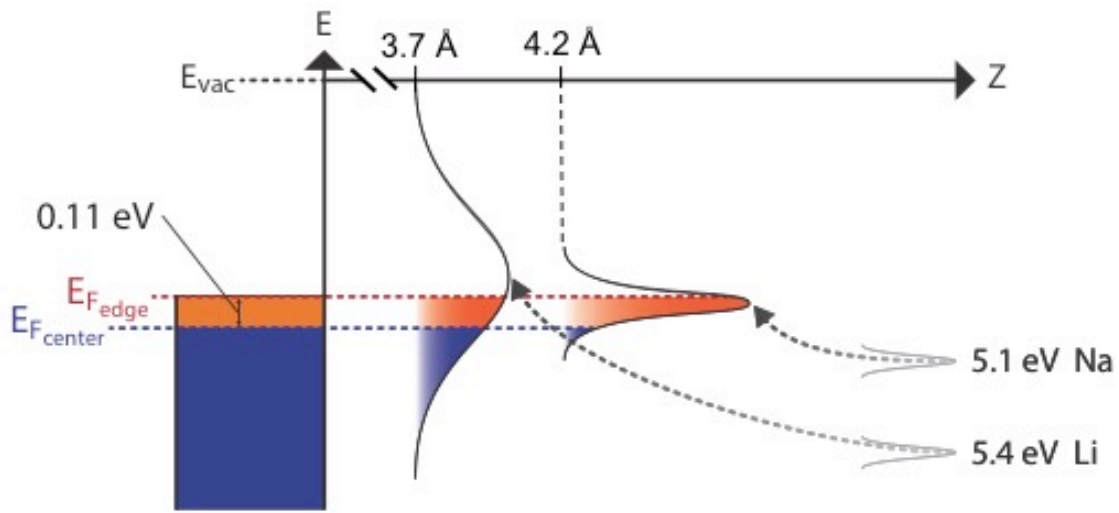

**Figure 5.** Schematic energy level diagram of the ion-solid system for Na and Li projectiles. Z refers to the atom-surface distance. The atomic *s* levels for Na and Li are shown at the right and labeled with their corresponding ionization energies, while the broadened and shifted ionization levels are shown at their freezing distances close to the surface (see text). The effective Fermi energies associated with the LEP above edge and center atoms are indicated by horizontal lines.